\documentstyle[12pt,epsfig]{article}
\begin{document}

\begin{flushright}
JYFL-6/98\\
hep-ph/9906438\\
\end{flushright}
\begin{centering}

\vspace{3cm}

{\bf Lower limits for $\langle E_{\rm T}\rangle$ and 
$\langle N_{\rm ch}\rangle$ from pQCD \& hydrodynamics at the central 
rapidity unit in central Au-Au collisions at RHIC\footnote{Presented
by K.J. Eskola in {\em Quark Matter '99}, 14 May, 1999, Torino, Italy,
in the session ``Last call for  predictions for RHIC.''  
For the transparencies, see \cite{ET}.
}}
\vspace{0.5cm}

	K.J. Eskola and K. Tuominen

\vspace{1cm}
{\em 
        Department of Physics, University of Jyv\"askyl\"a,\\
        P.O. Box 35, FIN-40351, Jyv\"askyl\"a, Finland\\
        kari.eskola@phys.jyu.fi, kimmo.tuominen@phys.jyu.fi
}
\vspace{1cm}
\begin{abstract}
Final state average transverse energy and charged particle
multiplicity at the central rapidity unit of central Au-Au collisions
at RHIC are studied within a framework of lowest order perturbative
QCD and Bjorken's hydrodynamical picture.  In particular, effects of
initial minijet production and the $pdV$ work during the subsequent
evolution of the QGP are investigated.  Nuclear shadowing effects are
included in a consistent manner, and the dependence of the results 
on the minijet transverse momentum cut-off parameter $p_0$ is shown.
\end{abstract}
\end{centering}
\vspace{1.0cm}

In heavy ion collisions at very high cms-energies the initial, very
early particle and transverse energy production at central rapidities
is expected to be dominated by multiple production of minijets,
i.e. gluons, quarks and antiquarks with transverse momenta $p_T\sim
1...2\, {\rm GeV} \gg \Lambda_{\rm QCD}$ \cite{BM}.  Assuming
independent multiple semi-hard parton-parton collisions, the average
transverse energy carried by the minijets produced with $p_{\rm T}\ge
p_0$ at the central rapidity window $\Delta y=1$ in a central (${\bf
b=0}$) $AA$-collision can be computed in the lowest order (LO)
perturbative QCD (pQCD) as \cite{EKL}

\begin{equation}
 E_{\rm T,pQCD}^{AA}(\sqrt s, p_T\ge p_0,\Delta y,{\bf 0}) 
= T_{AA}({\bf 0})
\int_{{p_0, \Delta y}} dp_Tdy_1dy_2 \frac{d\sigma}{dp_Tdy_1dy_2}p_T.
\end{equation}
The differential cross section above is that of each binary (LO)
parton collision, 
\begin{equation}
\frac{d\sigma}{dp_T^2dy_1dy_2} = 
        \sum_{{ijkl=}\atop{q,\bar q,g}}
        x_1f_{i/A}(x_1,Q) \, x_2f_{j/A}(x_2,Q)
        {d\hat\sigma\over d\hat t}^{ij\rightarrow kl} 
\end{equation}
where the rapidities of the outgoing partons are $y_1$ and $y_2$.  The
nuclear collision geometry is accounted for by the standard nuclear
overlap function $T_{AA}({\bf 0})\sim A^2/\pi R_A^2$. In order to
have at least one of the two outgoing partons in each minijet
collision in the rapidity window $\Delta y$, appropriate kinematical
cuts have to be made, see \cite{EKL,EKR,EK} for details. We use the
GRV94-LO parton distributions \cite{GRVLO}, and, to consistently
include both the $x$- {\it and} the $Q$-dependence of the nuclear
effects in the parton distributions (~$f_{i/A}(x,Q)\ne
f_{i/p}(x,Q)$~), we use the recent EKS98-parametrization \cite{EKS98}.
We emphasize that the results presented here are obtained by using
merely the lowest order pQCD but in order to approximate
the effects of the next-to-leading-order terms in the minijet cross
sections, we include $K$-factors 2.0 (1.5) for RHIC (LHC), together
with the scale choice $Q=p_{\rm T}$.

From saturation arguments \cite{EK}, and from requiring agreement with
inclusive pion spectra at central rapidities in $p\bar p$ collisions
at $\sqrt s = 200$ GeV, we expect $p_0\sim 1...2 $ GeV for Au-Au
collisions at RHIC. The few-GeV minijets are produced within a short
proper time $\tau_i\sim1/p_0\sim 0.2 ... 0.1$ fm/$c$, so they serve as
initial conditions for further evolution of the early QGP.  In
addition to the pQCD component, at RHIC we also expect a
non-negligible non-perturbative (soft) component in the initial
transverse energy production. For the simple estimates here, we take
the soft component directly from the measured average $E_{\rm T}$ at
$\eta\sim0$ at central Pb-Pb collisions at the SPS \cite{NA49}. The
{\it initial} transverse energy at $\tau=1/p_0$ thus becomes $E_{\rm
T}^i = E_{\rm T,pQCD}^{AA} + E_{\rm T}^{\rm NA49}$, as shown by the
dashed lines in Fig. 1a.  Dividing by the initial volume $V_i= \pi
R_A^2\tau_i\Delta y$, we get a Bjorken-estimate of the initial energy
density: $\epsilon_i = E_{\rm T}^i/V_i$. In a fully thermalized, 1+1
dimensional boost-invariant hydrodynamic system, the $pdV$ work causes
the energy density to decrease as $\epsilon\sim \tau^{-4/3}$ \cite{BJ}
and, especially, {\it the transverse energy to decrease} as $E_{\rm
T}\sim \tau ^{-1/3}$. We do not attempt to follow the system through a
phase transition here but simply decouple the system at $\epsilon_f =
1.5$ GeVfm$^{-3}$. The resulting $E_{\rm T}^{\rm final}$ represents a
lower limit of $\langle E_{\rm T}\rangle$ (mod the decrease in the
mixed/hadronic phase), and is plotted by the solid curves in Fig. 1a.
It should also be noted that when transverse expansion is included,
the loss of $E_{\rm T}$ is less than in the 1+1 dimensional case
considered above.

To get a corresponding estimate of the final state charged pion
multiplicity (Fig. 1b), we convert the initial energy density
$\epsilon_i$ into a temperature $T_i$, from which the initial rapidity
density of entropy $S_i$ can be computed.  For simplicity (inspite of
gluon dominance), let us assume full thermalization of gluons and
quarks here. In an isentropic boost-invariant 1+1 dimensional flow,
the rapidity density of entropy is conserved, so $S_i=S_f\approx
4\frac{3}{2}N_{\rm ch}$ \cite{EKR}, where $N_{\rm ch}$ is the final
state charged pion multiplicity in the central rapidity unit. In other
words, we obtain a lower limit of $N_{\rm ch}$ from entropy of the 
initial state.

To show the dependence on the transverse momentum cut-off, we plot the
lower limits for $\langle E_{\rm T}\rangle$ and $N_{\rm ch}$ as
functions of $p_0$ (the solid curves). The upper limit of $E_{\rm
T}^{\rm final}$ is the initial $E_{\rm T}^i$, so $p_0$ can be fixed
from some other arguments, the measured $\langle E_{\rm T}\rangle$
serves in principle also as a measure of thermalization. Our favorite
estimates for the lower limits can be read off from the figures at
$p_0\sim 1.5$ for RHIC and at $p_0\sim 2$ GeV for the LHC.

\begin{figure}[h]
\vspace{4.5cm}
\centerline{\hspace*{0.5cm} \epsfxsize=9.5cm\epsfbox{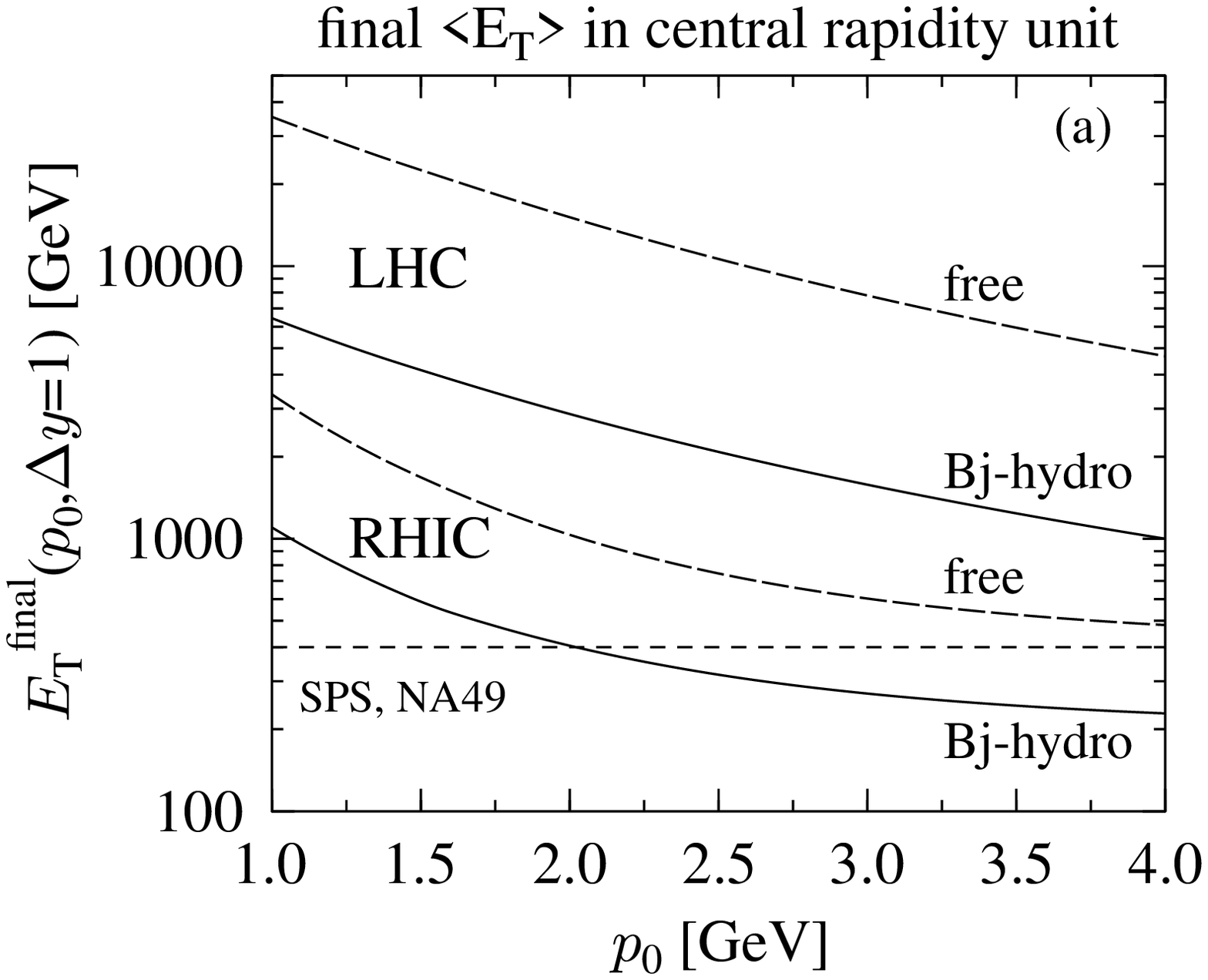}
\hspace*{-0.5cm}\epsfxsize=9.5cm\epsfbox{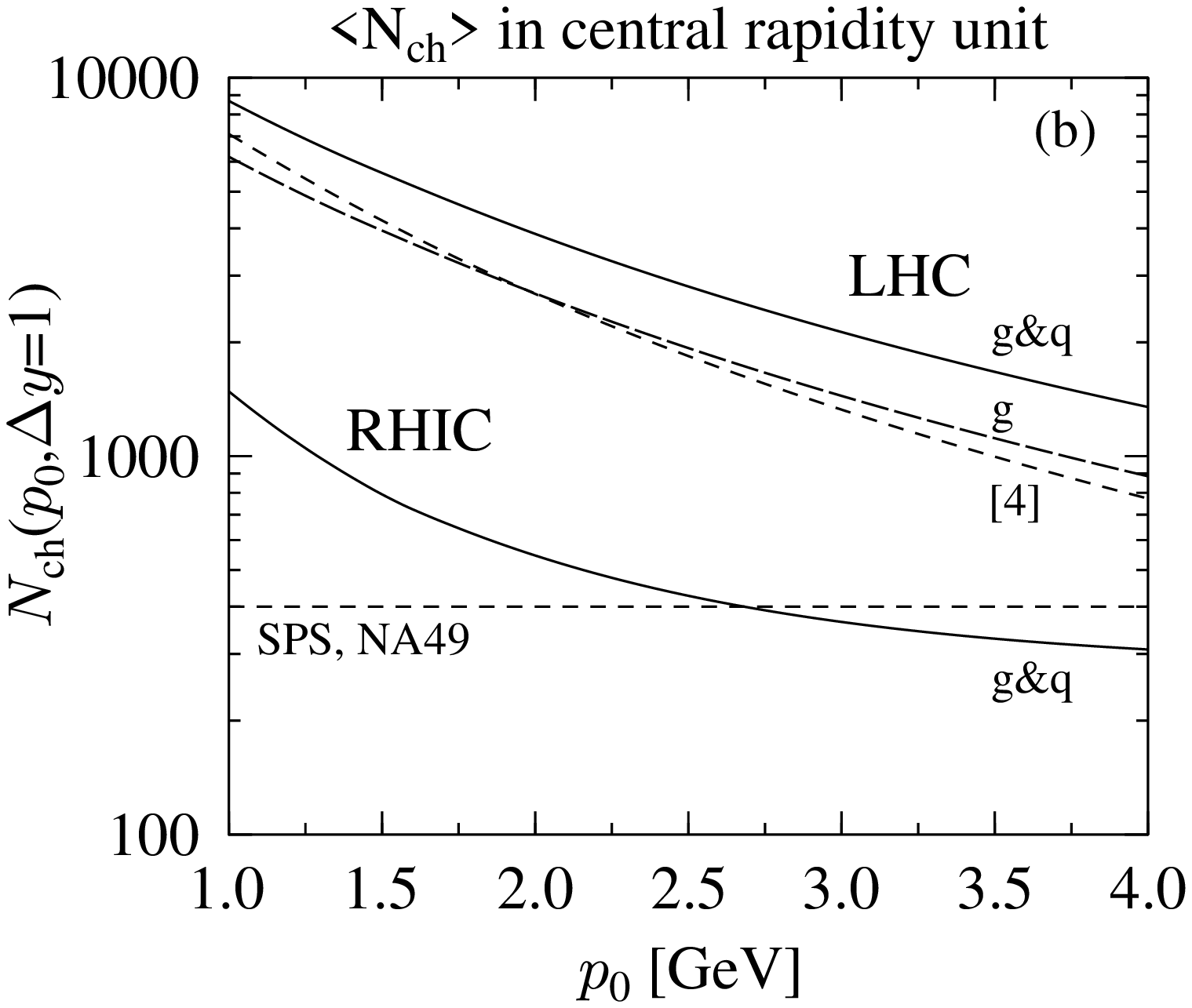}
}
\vspace{-6cm}
\label{DATA}
\end{figure}

\vspace{1cm}

\noindent{\bf Acknowledgements.} We thank K. Kajantie for useful 
dicussions, and the Academy of Finland for financial support.

\end{document}